\documentclass[preprint2]{aastex63}

\usepackage{CJK}
\usepackage{amsmath}
\usepackage{bm}
\usepackage{multirow}

\newcommand{\petit}{\texttt{petitRADTRANS}}

\newcommand{\pmn}{\texttt{DYNESTY}}

\received{\today}
\revised{}
\accepted{}
\submitjournal{ApJ}

\shorttitle{Early Massive Accretion}
\shortauthors{Wang}
\graphicspath{{./}{figures/}}

\begin{document}
\begin{CJK*}{UTF8}{gbsn}

\title{Early Accretion of Large Amounts of Solids for Directly-Imaged Exoplanets}

\correspondingauthor{Ji Wang}
\email{wang.12220@osu.edu}

\author[0000-0002-4361-8885]{Ji Wang (王吉)}
\affiliation{Department of Astronomy, The Ohio State University, 100 W 18th Ave, Columbus, OH 43210 USA}

\begin{abstract}

As the number of planetary mass objects (PMOs, $\lessapprox$13 M$_{\rm{Jupiter}}$) at wider separation ($\gtrapprox$10 AU) grows, there is emerging evidence that they form differently from their higher-mass brown-dwarf (BD) counterparts. Specifically, PMOs' atmospheres are often enriched by metals and show a large dispersion of metallicity, which is usually interpreted as a sign of solid accretion. {{As a first step toward a population-level study of the amount and timing of solid accretion, }}we analyze a sample of seven directly-imaged exoplanets with measured stellar and planetary chemical abundances (51 Eri b, $\beta$ Pic b, HIP 65426 b, HR 8799 c and e, AF Lep b, and YSES 1 c). Our analysis uses existing data of stellar and planetary atmospheric metallicities, and adopts a Bayesian framework that marginalizes the probabilities of disk conditions, formation locations, {{planetary interior structures}}, and accretion physics. We show that these PMOs accrete large amounts of solids {{regardless of whether they form via core accretion or disk instability}}. On average $\gtrapprox$50 M$_\oplus$ solids are accreted to enrich planet atmospheres. {{Individual planet accretes between 23.3 and 223.2 M$_\oplus$ of solid mass, more than 75\% of which is assumed to stay in the atmosphere and increase the observed metallicity.}} The result implies that the solid accretion process and therefore the planet formation process {{likely take place}} at an early stage {{($\lessapprox$2 Myr)}} when large amounts of solids are available in young {{massive}} protoplanetary disks. 

\end{abstract}



\section{Introduction}
\label{sec:intro}

As the population of directly imaged exoplanets continues to grow, there is an increasing interest in their formation origin and accretion history. There are two ways of forming planetary mass objects, a companion can form through gravitational instability (GI) in disks~\citep{Boss1997, Lodato2004, Durisen2007, Kratter2010}. Alternatively, a companion can form in a proto-planetary disk through core accretion~\citep[CA, ][]{Pollack1996, Bodenheimer2000, Hubickyj2005, Lissauer2009}. 

\begin{figure*}[th!]
\begin{tabular}{c}
\includegraphics[width=16.0cm]{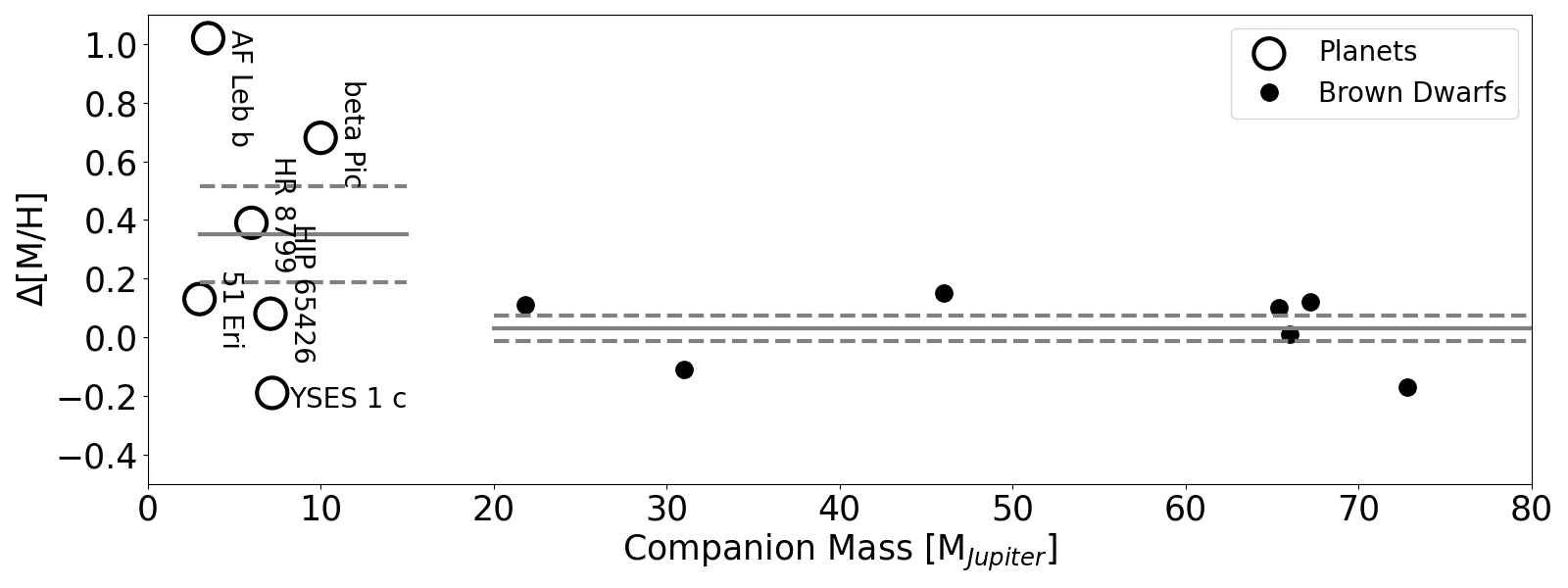}
\end{tabular}
\caption{{\bf{Directly-imaged exoplanets have super-stellar metallicities whereas higher-mass brown dwarf (BD) have stellar metallicities.}} The average relative metallicity ($\Delta$[M/H]) of directly-imaged exoplanets (open circles, HR 8799 c and e are shown as one data point for plotting clarity) is 0.35$\pm$0.16 dex (grey horizontal lines). $\Delta$[M/H] is the difference between the planet atmospheric metallicity and stellar metallicity. The error bar is the standard deviation divided by $\sqrt{N}$, where N=7 is the number of directly-imaged exoplanet systems. In comparison, the average metallicity of BDs (filled circles) is 0.03$\pm$0.04 dex (grey horizontal lines). The seven benchmark BDs are Gl 570 D, HD 3651 B~\citep{Line2015}, HD 19467 B~\citep{Maire2020}, HR 7672 B~\citep{Wang2022}, HD 4747 B~\citep{Xuan2022}, HD 33632 Ab~\citep{Hsu2024}, and YSES 1 b~\citep{Zhang2024}.  }
    \label{fig:planet_vs_bd}
\end{figure*}

There is suggestive evidence that planetary-mass companions and their higher-mass counterparts brown dwarfs (BDs) form through distinct mechanisms (Fig. \ref{fig:planet_vs_bd}). All directly-imaged exoplanets {{as a population has super-stellar metallicity}}, which suggests accretion of solids rich in Carbon (C) and Oxygen (O) that leads to super-stellar metallicities~\citep{Oberg2011}. In contrast, the higher-mass counterparts---BDs---show atmospheric metallicities that {{match stellar values within error bars, which is consistent with the}} multiple-star formation channel in which each component inherits a cloud/filament that is chemically identical.

Accretion history, regardless of forming through GI or CA, plays a critical role in the metal enrichment of directly imaged exoplanets. Previous studies show that a solid accretion mass of $\sim$100 M$_\oplus$ is needed to explain the atmospheric metal enrichment for $\beta$ Pic b~\citep{Nowak2020}, HR 8799 c~\citep{Wang2023} and e~\citep{Molliere2020}. This is puzzling because such a large mass reservoir may only be possible when the disk is young, which points to a very early planet formation ($<$1 Myr). Such an early planet formation is usually consistent with the GI channel, but recent works under the CA paradigm also show that that forming planets within 1 Myr is possible via early formation of solid cores, such as planetesimals by the streaming instability~\citep[e.g., ][]{Youdin2005,Li2021}, or embryos in self-gravitating disks~\citep{Baehr2022}, and then subsequent solid accretion~\citep{Ormel2010, Johansen2017, Baehr2023}.

It is therefore still {{debated}} via which dominant mechanism the directly-imaged exoplanets form, and {{it remains uncertain}} how much solid mass has been accreted {{and incorporated in the planet envelope}} depending on the formation mechanism and location. This paper aims to address these questions using a Bayesian inference framework. We organize the paper as follows. \S \ref{sec:observation} summarizes the collected data. \S \ref{sec:method} presents our model and the Bayesian framework. Main results are in \S \ref{sec:results}. Discussions are provided in \S \ref{sec:discussion}. A summary of the paper can be found in \S \ref{sec:summary}.

\section{Data}
\label{sec:observation}

\subsection{Data Acquisition}

The data for the seven exoplanets in six systems are presented in Table \ref{tab:data}. Our selection criteria require that the companion masses are below 13 M$_{\rm{Jupiter}}$ and their metallicity and C/O have been reported. While the choice of 13 M$_{\rm{Jupiter}}$ is arbitrary, changing the value will not change the conclusion that low-mass companions have super-stellar atmospheric metallicities and a higher dispersion in metallicities than their higher-mass counterparts. The gap of $\sim$10-30 M$_{\rm{Jupiter}}$ may be real and due to a BD desert as suggested in~\citet{Gratton2024}. 

{{
Fig. \ref{fig:planet_vs_bd} also shows BD companions around solar-type stars, i.e., benchmark BDs for which we can measure dynamical masses and directly compare metallicities with their primary stars~\citep{Line2015,Maire2020,Wang2022,Xuan2022}. {{In addition to being more massive than PMOs, the mass ratios of the benchmark BDs are $>$5\% (with the exception of Gl 570 D at 2\%). In comparison, mass ratios of PMOs are all $<$1\%. The contrast of mass ratios also indicates distinct origins of formation.}} While there are many more free-floating BDs with metallicity measurements~\citep[e.g., ][]{Manjavacas2019,Gonzales2020,Zhang2021b}, their inferred masses are less precise than dynamical masses and their formation origin is more difficult to assess without a comparison star. 
}}

{{For 51 Eri b, we use the stellar abundance from~\citet{Swastik2021} and the planetary metallicity from~\citet{BrownSevilla2022}. They use integral field unit (IFU) data in $Y$-, $J$- and $H$-band, and photometric data that cover $H$- through $M$-band and perform spectral retrieval using \petit~\citep{Molliere2019}. The metallicity reported in \petit\ assumes equilibrium chemistry and scales all metals {{(except for O)}}, so it is essentially [M/H]. The C/O is changed by changing O abundance. For $\beta$ Pic b, we assume solar values~\citep{Asplund2009} because of a lack of stellar abundance measurements. {{However, a recent chemical abundance measurement for HD 181327, another member of $\beta$ Pic moving group (BPMG), indicate BPMG C and O abundances can be super- ($\sim$0.2 dex) or sub-solar ($\sim$-0.2 dex) prior and post local thermal equilibrium (LTE) correction~\citep{Reggiani2024}. }} The planet metallicity is derived from a spectral retrieval analysis on a combined low-resolution (R$<$1000) data set of VLT/GRAVITY and Gemini-S/GPI~\citep{Nowak2020}. For HIP 65426 b, stellar metallicity is from~\citet{Swastik2021} and planet metallicity is derived using medium-resolution $K$-band SINFONI data and photometric data from $\sim$1 to 5 $\mu$m~\citep{Petrus2021}. They use a forward modeling approach in which self-consistent models are interpolated to fit the observed data. The best-fit [M/H] and C/O are then reported from their analysis. The HR 8799 stellar abundances are from~\citep{Wang2020}. For 8799 c, ~\citet{Wang2023} use high-resolution (R$\sim$35000) spectral data and archival low-to-medium-resolution data to measure both C and O abundances, i.e., [C/H] and [O/H]. The conversion between C and O abundances and [M/H] and C/O is given in \S \ref{sec:stage}. For HR 8799 e, [M/H] and C/O are measured based on low-to-medium-resolution data from Gemini-S/GPI, VLT/SPHERE, and VLT/GRAVITY~\citep{Molliere2020}. More recently, retrieval analysis on VLT/GRAVITY data (R=500) provides atmospheric abundances for AF Lep b~\citep{Balmer2024} and stellar abundances are from~\citet{Zhang2023}. For YSES 1 c, the data and analysis results are from VLT/CRIRES+ (R$\sim$100,000) by~\citep{Zhang2024}.  }}

\subsection{Caveats of the Data Set}

{There should be concerns of analyzing such a heterogeneous data set. However, the following arguments can alleviate the concern. First, planet atmospheric metallicities for all but one systems are derived using \petit\ with only one exception of HIP 65426, which uses \texttt{ForMoSa}}. The consistency in using the same atmospheric modeling code ensures the consistency of the analysis. This is evidenced by the consistent [M/H] and C/O ratios for HR 8799 c and e, which are derived independently using \petit\ and with two different data sets: one with high-resolution data~\citep{Wang2023} and the other one focused on low-to-medium resolution data~\citep{Molliere2020}. The cross-code consistency (i.e., \texttt{ForMoSa} vs. \petit) can be seen by the independent analysis on HIP 65426 b, although the uncertainties remain large~\citep{Wang2023b}.  

{{Moreover, PMO host stars and primary stars of the BD systems in Fig. \ref{fig:planet_vs_bd} have a systematic mass offset. The average mass of PMO host stars is 1.5 M$_\odot$ and 1.05 M$_\odot$ for primary stars of the BD systems. This systematic offset may complicate the interpretation of the difference between the PMO and the BD population. However, the number of PMOs around solar-mass stars are very limited. For example, AF Leb b and YSES 1 c have host stars with masses comparable with the average BD system primary stellar mass, and their relative metallicities represent the two extremes of the PMO sample, implying there is indeed an intrinsic difference between the PMO and BD population. On the other hand, benchmark BDs are rare around stars with masses around 1.5 M$_\odot$. Therefore, this sample of PMOs and benchmark BDs, despite its limitations, is the best available sample for the purpose of the analysis in this paper. 
}}

\subsection{Including Solar-System Planets}

In addition, we also include solar system gas giant planets, Jupiter and Saturn, for a comparative study (\S \ref{sec:jupiter_saturn}). The abundances for solar system planets are obtained from {{Table 2 in~\citet{Atreya2018}. We use [C/H] as a proxy of metallicity. Atmospheric C/H for Jupiter is measured with the Galileo Probe Mass Spectrometer down to 21 bar~\citep{Wong2004}; and C/H for Saturn is measured with the Cassini Composite Infrared Spectrometer based on CH$_4$ measurement~\citep{Fletcher2009}. We choose to use a solar C/O ratio---which remains to be consistent with state-of-the-art observation and modeling---because the C/O ratios of Jupiter and Saturn remain largely uncertain (see discussion in \S \ref{sec:jupiter_saturn}).  }}

\section{Method}
\label{sec:method}

{{We will first describe the mathematical framework to calculate the relative elemental abundance between a companion and its host star, as well as the conversion between metallicity and elemental abundances. We then proceed to describe our Bayesian modeling framework that links observables to inferred physical parameters such as accreted solid mass.  }}

\subsection{Setting the Stage}
\label{sec:stage}
The number {{of atoms for}} element X, we denote as $N_X$, can be calculated as follows:
\begin{equation}
    N_X = \frac{n_{X,s}}{\frac{f_{sg}}{1+f_{sg}}}M_s + \frac{n_{X,g}}{\frac{1}{1+f_{sg}}}M_g, 
\end{equation} 
where $n_{X,s}$ and $n_{X,g}$ are the number of X atoms per unit disk mass in solid and gas phase, $f_{sg}$ is {{the disk}} solid to gas ratio, and $M_s$ and $M_g$ are the masses in solid and gas phase, respectively. 

To calculate the mole ratio of the element X and H, for which we denote as X/H, for a planetary atmosphere that accretes some mass including $M_s$ and $M_g$. {{We assume the accreted gas stays in the atmosphere while the accreted solids contribute to both core growth and atmospheric metal enrichment. The ratio depends on ablation efficiency which will be discussed in \S \ref{sec:sampling}. As such, }} we have:

\begin{equation}
    \left(\frac{X}{H}\right)_{atm} = \frac{N_X}{N_H} = \frac{\frac{n_{X,s}}{f_{sg}}M_s + {n_{X,g}}M_g}{\frac{n_{H,s}}{f_{sg}}M_s + {n_{H,g}}M_g}. 
\end{equation}

Adopting the terminology in~\citet{Oberg2011}, the relative abundance $\alpha_X$ is:

\begin{equation}
\begin{split}
        \alpha_X & = \frac{\left(\frac{X}{H}\right)_{atm}}{\left(\frac{X}{H}\right)_{\ast}} \\
        & = \frac{\frac{\frac{n_{X,s}}{f_{sg}}M_s + {n_{X,g}}M_g}{\frac{n_{H,s}}{f_{sg}}M_s + {n_{H,g}}M_g}}{\frac{n_X}{n_H}} \\
        & = \frac{\frac{f_{X,s}}{f_{sg}}M_s + {f_{X,g}}M_g}{\frac{f_{H,s}}{f_{sg}}M_s + {f_{H,g}}M_g}, 
\end{split}
\end{equation}
where $n_X$ and $n_H$ are the number of X and H atoms per unit mass (the ratio of the two is by definition stellar mole ratio), $f_{X,s}$ and $f_{X,g}$ are the fraction of element X in solid and gas phase, and $f_{H,s}$ and $f_{H,g}$ are the fraction of hydrogen in solid and gas phase. We then assume the fraction of H in solid phase (i.e., $f_{H,s}$) is negligible because H$_2$, the most abundant materials, should be always in gas state under disk conditions. Given the assumption, the above equation becomes:

\begin{equation}
    \alpha_X = \frac{f_{X,s}}{f_{sg}}\frac{M_s}{M_g} + (1 - f_{X,s}),
    \label{eq:a_x}
\end{equation}
which is the same with Eq. 2 in~\citet{Oberg2011}. 

{{Since some literature reported [M/H] (M is total metal content) and C/O~\citep[e.g., ][]{Nowak2020, Molliere2020, Petrus2021, BrownSevilla2022} instead of $\alpha_X$ for C and O, we provide the conversion between the two conventions. Assuming metals in planet atmospheres {{scale with C abundance with the exception of O}}, we have the following equation, starting in a similar form as Eq. \ref{eq:a_x}:

\begin{equation}
\begin{split}
    \log\alpha_M & = \log\left(\frac{\left(\frac{M}{H}\right)_{P}}{\left(\frac{M}{H}\right)_{\ast}}\right) \\
    & = \log\left(\frac{\left(\frac{M}{H}\right)_{P} / \left(\frac{M}{H}\right)_{\odot}}{\left(\frac{M}{H}\right)_{\ast} / \left(\frac{M}{H}\right)_{\odot}}\right) \\
    & = [M/H]_P - [M/H]_\ast  \\
    & = \log\left(\frac{\left(\frac{C+O}{H}\right)_P}{\left(\frac{C+O}{H}\right)_\ast}\right) \\
    & = \log\left(\frac{(1 + (\frac{C}{O})_P)(\frac{O}{H})_P}{(1 + (\frac{C}{O})_\ast)(\frac{O}{H})_\ast}\right) \\
    & = \log\left(\frac{1 + (\frac{C}{O})_P}{1 + (\frac{C}{O})_\ast}\right) + \log\left(\frac{(\frac{O}{H})_P}{(\frac{O}{H})_\ast}\right) \\
    & = \log\left(\frac{1 + (\frac{C}{O})_P}{1 + (\frac{C}{O})_\ast}\right) + \log\alpha_O.
    \label{eq:m_h}
\end{split}
\end{equation}

Eq. \ref{eq:m_h} gives the conversion between $\alpha_X$ (X is C or O) and reported properties such as [M/H]$_P$, [M/H]$_\ast$, {{i.e., atmospheric metallicity for the planet and the star,}} and {{their}} C/O ratios. 

{{A major caveat of the assumption that all metals abundances scale with C is that N, which is another major volatile element, may not scale with C for the same reason as the changing C/O as a function of distance to the star. N abundance and C/N ratio will depend on the physical condition of the disk such as temperature and pressure~\citep{Turrini2021}. As a result, N is predicted as an additional tracer for planet formation in the JWST era~\citep{Ohno2023}. However, without a {{reliable}} N measurement for our sample, we will assume that N scales with C and Eq. \ref{eq:m_h} can be easily adapted to a case with a non-scaling C/N ratio.   }}

}}

\subsection{Model Input and Output}
\label{sec:input_output}

We adopt the values in Table \ref{tab:co_carriers} for major C and O carriers in a proto-planetary disk. The planet atmospheric chemical composition, which is parameterized as $\alpha_C$, $\alpha_O$, and C/O, is determined by (1) where the planet forms; (2) how it forms; and (3) the amount of solid accretion. These parameters are compared to the observed values to infer the formation location, mechanism, and the accretion history. In addition, values in Table \ref{tab:co_carriers} are scaled to matched the observed host star values. {{Table \ref{tab:input_output} and \S \ref{sec:stellar_mh_co} provide more details on the scaling: each C and O carrier in Table \ref{tab:co_carriers} is given a different scaling factor to account for the non-solar C and O abundances. }}

Table \ref{tab:input_output} {{also}} summarizes the input and output of our model. There are {{five}} input parameters: {{stellar C and O abundances denoted as $\xi_C$ and $\xi_O$,}} and C/O, planetary relative metallicity to the host star, i.e., [M/H]$_p$ - [M/H]$_\ast$, and planetary atmospheric C/O. The parameters to be inferred are: {{scaling factors for C and O carriers in Table \ref{tab:co_carriers}}}, planetary mass, mass of accreted solids, solid-to-gas ratio $f_{sg}$, formation mechanism, the disk temperature where a planet forms which indicates the planet formation location, initial core mass, and ablation fraction. The priors for these parameters are given and justified in \S \ref{sec:sampling}. 

\subsection{{{Stellar Metallicity and C/O}}}
\label{sec:stellar_mh_co}

Table \ref{tab:input_output} also provides equations that connect different parameters. For example, the scaling factors for stellar C and O abundances are used together with the C and O carrier abundances in Table \ref{tab:co_carriers} to match with stellar $\xi_C$, $\xi_O$ and C/O.{{ Specifically, $\xi_C$ is calculated by $\log{\sum n_{C_i}\cdot \epsilon_{i}}$, where $\epsilon_i$ is the scaling factor for C carriers such as CO, CO$_2$, and C grains, and $n_{C_i}$ is their abundances in Table \ref{tab:co_carriers}. $\xi_O$ is similarly calculated with $n_{O_i}$ being abundances for O carriers in Table \ref{tab:co_carriers}, and C/O is obtained using ${\sum n_{C_i}\cdot \epsilon_{i}}/{\sum n_{O_i}\cdot \epsilon_{i}}$. }}

\subsection{{{Planet Metallicity and C/O}}}

The planetary mass ($M_P$), mass of accreted solids ($M_s^\prime$), initial core mass ($M_{\rm{core}}$), ablation fraction ($\eta$), solid-to-gas ratio $f_{sg}$, and the disk temperature (T$_d$) are used for calculating planet relative metallicity and C/O. 

{{Using {{Carbon C}} as an example, for a given T$_d$, we can determine the phase (solid vs. gas) for each C carrier in Table \ref{tab:co_carriers}. The fraction of C in solid phase, $f_{C,s}$ in Eq. \ref{eq:a_x} is then calculated. In order for $\alpha_C$ to be determined, The mass of accreted solids that contribute to the atmospheric metallicity, i.e., $M_s$, is $M_s = M_s^\prime \times \eta$. The gas mass $M_g$ is calculated as $M_g = M_P - M_s^\prime - M_{\rm{core}}$. }}


With $f_{C,s}$, $f_{sg}$, $M_s$, $M_g$, $M_{\rm{core}}$, and $\eta$, C abundance $\alpha_C$ is determined via Eq. \ref{eq:a_x} and associated equations in Table \ref{tab:input_output}. O abundance $\alpha_O$ is determined in a similar way. {{To compare to the observed planet relative metallicity to the host star, i.e., [M/H]$_p$ - [M/H]$_\ast$, we use Eq. \ref{eq:m_h}.}} To compare to the observed C/O, we use $\alpha_C/\alpha_O\times C/O_\ast$, where $C/O_\ast$ is ${\sum n_{C_i}\cdot \epsilon_{i}}/{\sum n_{O_i}\cdot \epsilon_{i}}$ as noted in Table \ref{tab:input_output}.  

\subsection{Bayesian Sampling and Model Selection}
\label{sec:sampling}

We compare the model output to the input data (listed in Table \ref{tab:data}) in order to infer parameters such as accreted solid mass (listed in full in Table \ref{tab:input_output}. The priors for inferred parameters are given in Table \ref{tab:prior}. {{The five scaling factors have a flat uniform prior from 0.5 to 2.0. Planet mass priors are set based on values in Table \ref{tab:data} from the cited literature. The prior for the accreted solid mass is a log-uniform function from $10^0$ to $10^4$ M$_\oplus$. {{We use a nonphysically high upper limit for accreted solid mass to prevent the posterior sample to pile up at the upper bound of the chosen prior. The upper limit will not affect our result because all inferred solid masses are well below the limit. }}

{{The log-normal prior for $\log{f_{sg}}$ is centered at -2.0 with a 1-$\sigma$ dispersion of 0.5. This corresponds to a 95\% credible interval for ${f_{sg}}$ from 0.1\% to 10\%.}} The range covers the higher end of $f_{sg}$ from~\citep{Ansdell2016} and lower end of $f_{sg}$ due to underestimation of CO gas~\citep{Anderson2022} and the neglecting of CO depletion~\citep{Cleeves2016, Yu2016, Bosman2018, Zhang2021}. 

Disk temperature prior is set to be log-uniform from $10^{1.0}$ to $10^{2.5}$ K, covering temperature range that is twice as cold as the CO condensation temperature and twice as hot as the $H_2O$ evaporation temperature. {{We note that the evaporation temperatures listed in Table \ref{tab:co_carriers} depend on the assumed binding energy~\citep{Cuppen2017} and number density~\citep{Hollenbach2009}. For example, the range of disk mid-plane gas density (10$^8$-10$^{12}$ cm$^{-3}$) corresponds to $\sim$15\% change of the evaporation temperature. However, this change will not significantly affect our conclusion as long as our logarithmic prior covers the evaporation range for the considered species and each temperature range is roughly equal in the log-uniform space.    }}

Initial core mass prior is set to be log-uniform from $10^{-1.0}$ to $10^{2.0}$ M$_\oplus$. The choice of the prior roughly covers three scenarios: GI in which the core mass is negligible (0.1 to 1.0 M$_\oplus$), CA with a light (1 to 10 M$_\oplus$) and heavy core (10 to 100 M$_\oplus$) as shown in~\citet{Drazkowska2023}. 

The prior for the ablation fraction is uniform from 50\% to 100\%. These values are consistent with investigation on ablation efficiency of icy and carbonaceous planetesimals of sizes from 10 m to 1 km~\citep{Pinhas2016}. The ablation fraction for smaller pebbles should be on the higher end close to 100\% according to Fig. 4 in~\citet{Pinhas2016}. }}

\begin{figure*}[th!]
\begin{tabular}{c}
\includegraphics[width=16.0cm]{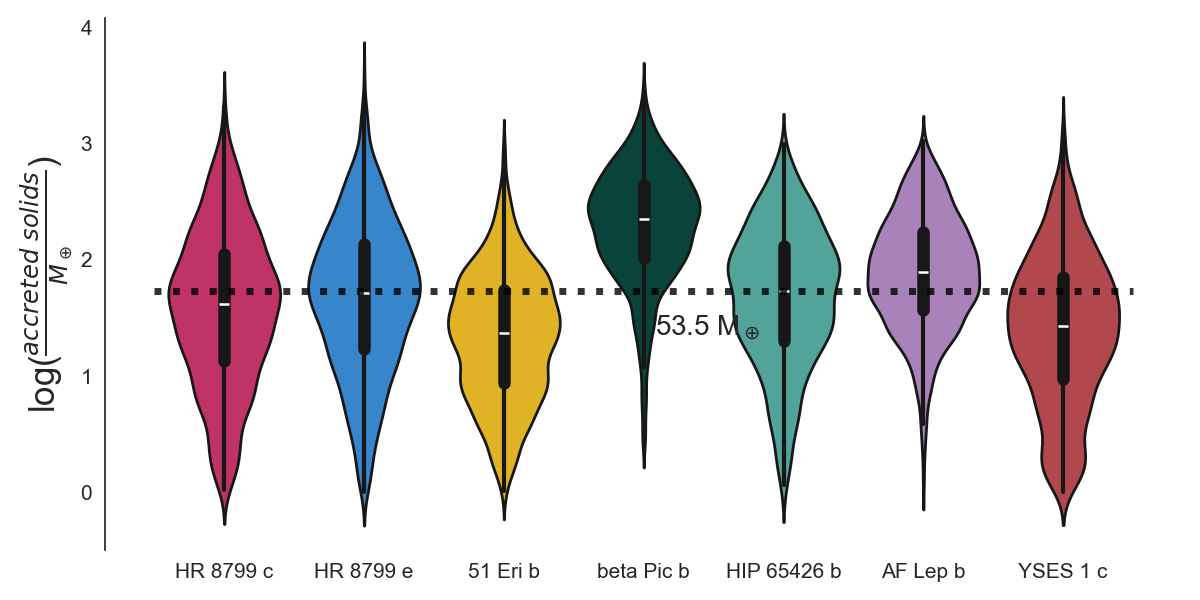}
\end{tabular}
\caption{{\bf{Directly-imaged exoplanets accrete on average $\gtrapprox$50 M$_\oplus$ solids to enrich their atmospheres.}} {{Each violin shape shows the inferred accretion solid mass distribution (vertical axis) for a directly-imaged exoplanet. The median of the overall distribution is represented by the white point, and the 25\% and 75\% percentiles are represented by the thick line. An average value of 53.5 M$_\oplus$ is given as the horizontal dotted line.}}  }
    \label{fig:accreted_mass}
\end{figure*} 

The likelihood function we use is: ${-\frac{1}{2}\displaystyle\sum_{i=1}^{n}{(\mathcal{D}_i-\mathcal{M}_i)^2/\mathcal{E}_i^2}} - \frac{1}{2}\displaystyle\sum_{i=1}^{n}{\ln(2\pi \mathcal{E}_i^2)}$, where subscript $i$ is parameter index, $D$ is data, $M$ is model, and $\epsilon$ is error. We used \pmn~\citep{Speagle2020} to sample posteriors and calculate evidence. Examples of retrieving accretion mass can be found in~\ref{sec:diagnostic}. 

{{Convergence is defined as three consecutive runs which (1) are not ordered in their log(Z) results (Z is the evidence or marginal likelihood), and (2) the consecutive log(Z) error bars overlap. We also try different setups for parameters that affect convergence to test the sensitivity of posterior samples to these parameters. The parameters that we test include live points at 1000, 2000, and 5000; evidence tolerance at 0.05, 0.1, and 0.5. We find that our results are not sensitive to the choice of these parameters. }}

\section{Results}
\label{sec:results}

\subsection{Order of Magnitude Estimation}
\label{sec:oom}

{{The above Bayesian framework is useful in marginalizing uncertainties of modeling parameters and estimating the final uncertainty of accreted solid mass. However, a much more simpler order-of-magnitude calculation can get us to the ballpark range and offers a much more clear physical insight than through the opaque Bayesian curtain. Below, we will provide such order-of-magnitude estimations of the accreted solid mass under two different conditions. 

First, we assume formation beyond the CO iceline. This is the simplest case because both $f_{C,s}$ and $f_{O,s}$ is unity. We can use Eq. \ref{eq:a_x} to estimate $M_s$ assuming reasonable values for other parameters. We choose $\log\alpha_M$ to be 0.34, the median value of relative metallicity in Table \ref{tab:data}. The choice of C or O only changes results by a factor of 2, within an order of magnitude. Similarly, we can assume $M_g$ to be 5 M$_{\rm{Jupiter}}$, which is $\sim$1500 M$_\oplus$, and $f_{sg}$ to be 1\%. Plugging these values, we have $M_s$ to be $\sim$30M$_\oplus$. Since $M_s = M_s^\prime \times \eta$, the accreted solid mass $M_s^\prime$ is therefore $\sim$40M$_\oplus$ assuming $\eta=0.75$, the median our chosen prior. 

Second, we assume another extreme case in which a planet forms within the water iceline. In this case, $f_{C,s}$ and $f_{O,s}$ is around 20-30\%, for which we use 25\%. If all other parameters are the same as the previous case, $M_s^\prime$ is $\sim$100 M$_\oplus$. The required accreted solid mass to reach the same relative metallicity is higher for formation within water iceline than beyond the CO iceline. This is also seen in the discussion of solar system gas giant planets in \S \ref{sec:jupiter_saturn}. }}

\subsection{Mass of Accreted Solids}
\label{sec:accreted_mass}

Mass of Accreted Solids is inferred in our Bayesian modeling (\S \ref{sec:method}) ans is presented in Table \ref{tab:jupiter_saturn}. On average, a planet accreted $\gtrapprox$50 M$_\oplus$ solids, with a noticeable outlier of $\beta$ Pic b which accretes $\sim$200 M$_\oplus$ solids (Fig. \ref{fig:accreted_mass}). For a typical 5-$M_{Jupiter}$ planet in our sample, 50 M$_\oplus$ corresponds to 3\% fractional mass for metals mainly in the form of volatile ice. This is roughly consistent with 0-20\% volatile fractions for simulations for gas giant planets~\citep{Mordasini2016}.  

{{Our approach is agnostic about formation mechanism (i.e., CA vs. GI) because the prior initial core mass ranges from essential no core (0.1 M$_\oplus$) to a massive core (100 M$_\oplus$)}}. {{Forming directly-imaged planets via the GI channel is a viable way, ~\citet{Boss2023} shows that GI can form companions with masses ranging from sub-Jupiter to low-mass star in 2,000 years. However, longer GI simulations often form companions with masses over 10 $M_{Jupiter}$~\citep{Zhu2012} which are often more massive than the observed planets.}} In addition, GI is unlikely to explain the observed properties for some of the planets in our sample, e.g., the orbital period commensurability for HR 8799 planets~\citep{Konopacky2016} and the low luminosity for 51 Eri b~\citep{Macintosh2015}.

\begin{figure}[h!]
\begin{tabular}{c}
\hspace*{-0.9cm}
\includegraphics[width=0.99\linewidth]{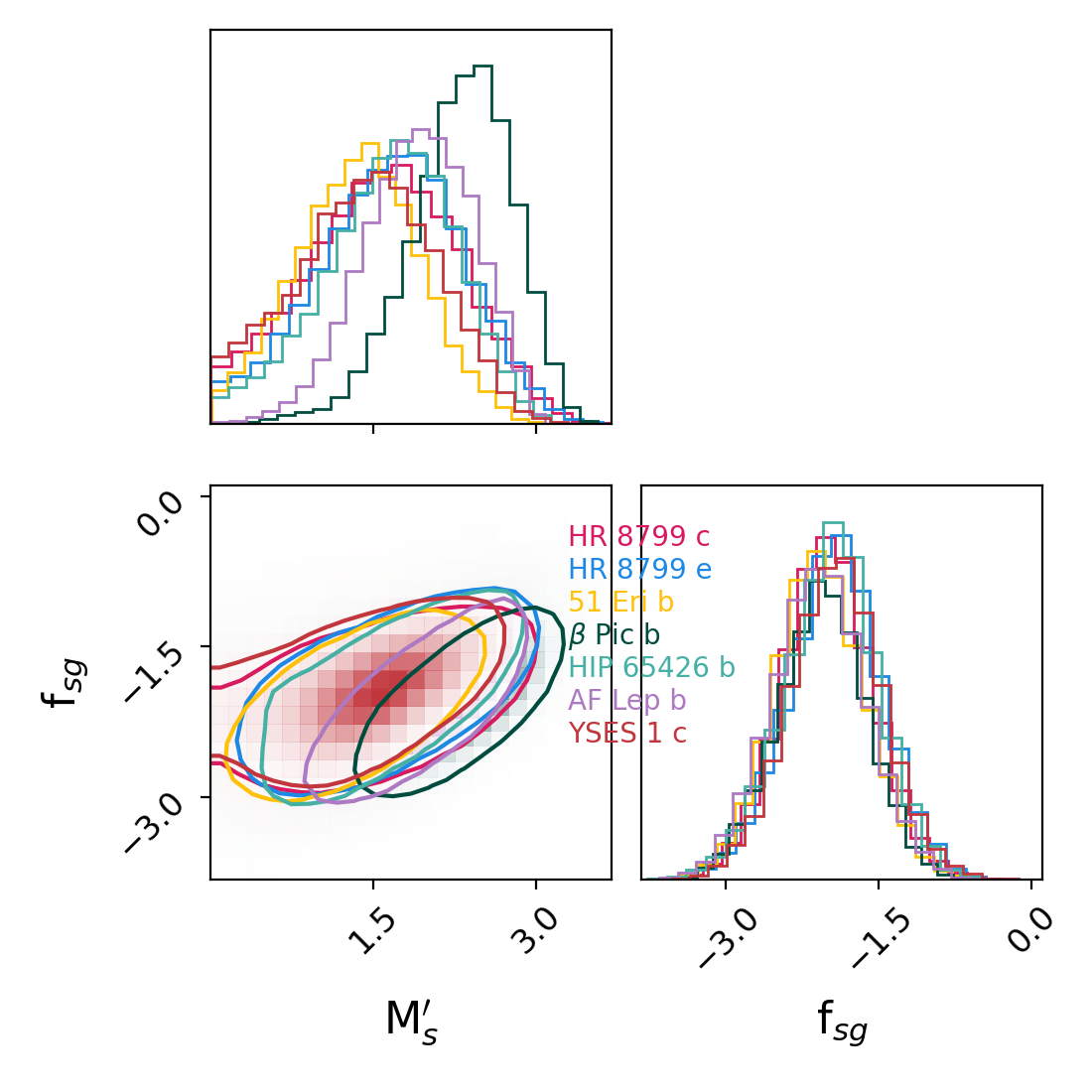}
\end{tabular}
\caption{{\bf{Inferred accreted solid mass positively correlates with assumed solid-to-gas ratio for all seven directly-imaged systems (marked by colors).}} The correlation ranges from 0.62 to 0.93 dex increase in solid mass for every 1 dex increase in $f_{sg}$. The inferred accreted solid mass is sensitive to the choice of $f_{sg}$. See more discussions on the correlation in \S \ref{sec:accreted_mass} and \S \ref{sec:beta_pic}.   }
    \label{fig:correlation}
\end{figure} 

We note that the inferred accreted solid mass ($M_s^\prime$) is positively correlated with the assumed $f_{sg}$ (shown in Fig. \ref{fig:correlation}) as there is a degeneracy between $M_s^\prime$ and $f_{sg}$ in Eq. \ref{eq:a_x}. For the seven directly-imaged planets, the correlation ranges from 0.62 to 0.93 dex in $M_s^\prime$ for every 1 dex increase in $f_{sg}$. {{Similarly, $M_s$ and $f_{sg}$ are positively correlated because $M_s$ and $M_s^\prime$ are positively correlated by an ablation efficiency $\eta$ (See Table \ref{tab:input_output}). }}
Therefore, the inferred accreted solid mass is sensitive to the choice of $f_{sg}$ prior. Our $f_{sg}$ log-prior is centered at 1\% covering a wider range from 0.1\% to 10\% at 2-$\sigma$ level, the retrieved mean value of the accreted mass should correspond to the $f_{sg}$ value of 1\% and with a large uncertainty due to the large prior range.     

{{Related to the positive correlation between $M_s^\prime$ and $f_{sg}$, two assumptions on solid-to-gas ratio $f_{sg}$ at the planet location can lead to different inferred accretion solid masses. In our framework, we assume that $f_{sg}$ is independent of core mass because of the resupply of solids due to migration~\citep{Alibert2004} or pebble accretion~\citep{Bosman2023}. This is evident by the lack of correlation between core mass and accreted solid mass in Fig. \ref{fig:corner_solar_j} and \ref{fig:corner_solar_s}. Alternatively, $f_{sg}$ can be negatively correlated with core mass because the growth of a planetary core depletes the solids in its feeding zone~\citep{Pollack1996}. In this scenario, local $f_{sg}$ is generally lower and therefore leads to lower inferred accreted solid mass because of the positive correlation between $M_s^\prime$ and $f_{sg}$.    
}}

\subsection{Timing of the Solid Accretion}
\label{sec:accreted_time}

The large amount of accreted solids is difficult to be explained by the solid mass budget of a typical disk. At 1\% {{disk-to-star}} mass ratio with a solar metallicity, the solid mass budget is $\sim$60 M$_\oplus$, which is barely sufficient for single-planetary systems even assuming close to 100\% accretion efficiency, not to mention the multi-planetary systems such as HR 8799 and $\beta$ Pic.  

Since disk mass and the solid content therein decrease with time, $\sim$50\% of $\sim$$10^5$ yr-old young Class 0 disks~\citep{Tychoniec2020} show dust mass at $\sim$100 M$_\oplus$. By the age reaches $\sim$1 Myr, at most 25\% of systems have  solid content at $\sim$30 M$_\oplus$~\citep{Drazkowska2023}, which is insufficient to {{explain the solid content of}} the directly-imaged exoplanets in our sample. Therefore, the formation of these planets and the necessary solid accretion {{is likely}} to take place within 1 Myr. 


{{However, there are caveats in the above arguments. {{First, extremely early planet formation~\citep[e.g., ][]{SeguraCox2020} and disk mass can be supplied and replenished through streamers in the Class-0 or I envelope phase~\citep{Visser2009} and therefore do not need to resort to the conventional disk mass reservoir. Second, }}the drastic decrease of solid content with time as inferred from~\citet{Tychoniec2020} is for mm-cm solids that ALMA observations are sensitive to. Therefore, the drastic decrease can be interpreted as either a sharp decrease of total solid mass, or a quick growth from mm-cm solids to larger solids such as pebbles and/or planetesimals. The latter may be supported by recent studies of pebble drift rate~\citep[e.g.,][]{Bosman2023} and solar system iron meteorites~\citep{Kleine2005, Schersten2006}. 

From an observational perspective, signs of planet formation (e.g., annular features) have been observed at $<$0.5 Myr~\citep{SeguraCox2020} to support the early planet formation within 1 Myr. However, the annular features for a young class-I object as seen by~\citet{SeguraCox2020} may be attributed to other effects than planet formation such as ice sintering~\citep{Okuzumi2016}, magnetically dead zones~\citep{Ruge2016}, and zonal flows~\citep{Uribe2011}. Even if the features are due to dusts that are forming planets, there should be some time lag between an annulus appearing and a Jupiter-mass planet forming. 

From an simulation perspective, to facilitate early planet formation requires a high disk-to-star mass ratio~\citep[e.g., $\sim$5\%, ][]{Savvidou2023,Baehr2023}. {{Indeed, $\sim$10-20\% disk-to-star mass ratio has been observed for many systems with disk features such as rings and spiral arms that are indicative of planet formation~\citep{Powell2019, Veronesi2021, Lodato2023}. Seven out of ten systems are younger than 2 Myrs. This again points to the early timing for planet formation.}} Taken together, the timing of the solid accretion can be as early as $\sim${{1-2}} Myrs if available solids decrease drastically with time, but can also last longer if solids growth is more quickly than $\sim${{2}} Myrs. }}


\section{Discussion}
\label{sec:discussion}

\subsection{Jupiter and Saturn}
\label{sec:jupiter_saturn}


We apply our inference framework to the solar system gas giant planets, Jupiter and Saturn, to check if the inferred mass of solid accretion agrees with previous studies.  

To ensure a fair comparison, we sum up the accreted mass and the initial core mass to calculate the total metal mass. The total metal mass in Jupiter is $ 34.6^{+ 38.9}_{- 21.1}$ M$_\oplus$ {{(reporting 16, 50, and 84 percentile of posterior distribution)}}, consistent with values in~\citet{Guillot1999,Thorngren2016} within error bars. Our lower limit is comparable to theirs, but our upper limit is considerably higher, possibly due to our large prior range.  

Upon further inspection, the inferred accreted mass depends on the formation location with a 2:1 preference to beyond the CO iceline (Fig. \ref{fig:corner_solar_j} {{with red vertical lines marking condensation temperatures for CO and H$_2$O}}). This is consistent with previous claim of formation beyond N$_2$ iceline~\citep{Oberg2019}. The measured low Oxygen abundance ($<$5x solar) from the Juno mission~\citep{Li2020} and the associated thermochemical modeling~\citep{Cavalie2023} also supports the formation location that is outside the CO iceline. The total metal mass in Jupiter if assuming formation beyond the CO iceline is $ 27.6^{+ 25.3}_{- 16.6}$ M$_\oplus$, more consistent with literature values~\citep{Guillot1999,Thorngren2016,Militzer2022}.

{{Despite the consistency, however, we note that formation beyond the CO iceline is not supported by meteoritic evidence~\citep{Morbidelli2016,Kruijer2017}. Considering N abundance and the N-bearing specie icelines in our model may resolve the issue, but this is currently beyond the scope of our model and this paper.  }}



Saturn, similarly showing a multi-modal distribution of inferred accretion mass depending on formation locations. The total metal mass within Saturn is $ 13.9^{+ 11.6}_{-  6.3}$ M$_\oplus$ if forming beyond the CO iceline (the preferred formation location from our inference, see also Fig. \ref{fig:corner_solar_s}). This formation location is consistent with Jupiter forming outside the CO iceline since Saturn is located outside Jupiter. Given that the Oxygen abundance for Saturn is largely uncertain~\citep{Atreya2018}, we assume a solar C/O and note that the actual accreted solid mass may be more uncertain and have a systematic shift from the inferred value. However, our results are broadly consistent within 1-$\sigma$ when compared to previous works~\citep{Guillot1999,Thorngren2016,Mankovich2021}.

{{While the total metal content of Saturn is lower than that of Jupiter, Saturn may have a higher core mass than Jupiter~\citep{Saumon2004}. However, the exact core mass and metal content therein depend on models~\citep{Nettelmann2012} and the spatial extent of a core~\citep{Ni2019}. }}


{{Since the next favorable formation location is within the H$_2$O iceline, we discuss the this scenario as follows. }}Formation of solar-system gas giant planets within H$_2$O iceline required higher accretion mass. This has also been seen in the order-of-magnitude calculation in \S \ref{sec:oom}. If that is the case, then the inferred total metal mass is $ 47.6^{+ 48.2}_{- 28.2}$ M$_\oplus$ for Jupiter. This higher than literature value can be alleviated by invoking the pebble evaporation mechanism~\citep{Booth2017, Mousis2019, Schneider2021,Bitsch2023}. Evaporating pebbles, especially in the inner disk where temperature is high, can enhance the local gas metallicity and therefore enrich the planet atmospheric metallicity during the gas accretion process. Combining pebble evaporation and solid accretion to explain the metallicity enrichment will reduce the mass requirement for solid accretion alone. 




\subsection{Atmospheric and Bulk Metallicity}
\label{sec:atm_bulk_z}

{{

It is well known that the metallicity distribution within Jupiter and Saturn is not uniform. There are two caveats in using the measured atmospheric metallicity to inform accretion history. First, a fuzzy core may enrich the atmospheric metallicity~\citep{Wilson2012, Madhusudhan2017}. However, this scenario requires a diluted core and a large core mass to account for the inferred $\sim$50 M$_\oplus$ in the atmosphere. This is inconsistent with our core assumption at $\sim$20 M$_\oplus$. 

Second, the required amount of solids in the atmosphere is reduced if the atmosphere layer, which is assumed to convective and isolated from the core in our model, is thinner. Indeed, ~\citet{Shibata2022} show that the required solid accretion is $\sim$4 M$_\oplus$ if Jupiter's outermost fully mixed layer is less than 20\% of Jupiter mass. Scaling up the value for a gas giant planet with a clearly-define core and a fully convective envelop that dominates the mass fraction, the requirement for solid accretion is $\sim$20 M$_\oplus$, consistent with our inferred value (Table \ref{tab:jupiter_saturn}). 

While current Jupiter and Saturn observations and models favor a fuzzy core that extends to $\sim$50\% of planet radius~\citep{Ni2019,Mankovich2021}, the core may be more confined and the convective envelope is thicker right after the planet forms, and it may take 100 Myr to 1 Gyr to form a fuzzy core~\citep{Helled2022}. Since all planets in our sample are much younger than 100 Myr, the assumption of a thick convective envelop that dominates the planet mass budget is not unreasonable.  




}}

\subsection{Comparing to Previous Results}
\label{sec:comp}

\subsubsection{$\beta$ Pic b}
\label{sec:beta_pic}

~\citet{Nowak2020} adopt a similar approach to investigate solid accretion history for $\beta$ Pic b. The major differences between their approach and this work are (1) we consider both [M/H] and C/O while they consider only C/O; and (2) we use a Bayesian inference framework taking fully into consideration of measurement uncertainties. While ~\citet{Nowak2020} conclude that $\beta$ Pic b is unlikely to form via the gravitational instability mechanism because of unusually long time for the pre-collapse phase, or an extremely efficient accretion (with an accretion rate of 4000  M$_\oplus$ per Myr), we note that a pebble influx rate as high as 1000  M$_\oplus$ per Myr has been observationally supported~\citep{Bosman2023}. {{In addition, the inferred mass for solid accretion of $\sim$300 M$_\oplus$ would also suggest a massive disk which is conducive for gravitational instability.}} Therefore, forming $\beta$ Pic b through gravitational instability is still a plausible channel. 

Core accretion scenario for $\beta$ Pic b has also been considered in~\citet{Nowak2020}. They calculate that 80-150 M$_\oplus$ needs to be accreted to reach the 1-$\sigma$ upper limit of C/O measurement. In comparison, this work infer $294.6^{+315.2}_{-156.8}$ M$_\oplus$ needs to be accreted in order to explain the observed C/O and planet metallicity. 


While roughly consistent with previous estimation, several factors that may be responsible for the difference include (1) them using the 1-$\sigma$ upper limit of C/O whereas us using the median value; (2) whether or not formation location is marginalized; and (3) whether C/O is used alone or both [M/H] and C/O are used in the inference. We also note that [M/H] for $\beta$ Pic b has a less preferred value at -0.53$^{+0.28}_{-0.34}$ in~\citet{Nowak2020}, which would significantly reduce the inferred mass for solid accretion. This highlights the issue that our inference is highly sensitive to systematics in measurements of abundances and abundance ratios.        


\subsubsection{HR 8799 c and e}

HR 8799 c~\citep{Wang2023} and e~\citep{Molliere2020} are found to be metal enriched. It is estimated that a solid mass of $\sim$100 M$_\oplus$ needs to be accreted~\citep{Molliere2020}. This is roughly consistent within large uncertainties with our values at $\sim$50 M$_\oplus$ in this work (see Fig. \ref{fig:accreted_mass} and Table \ref{tab:jupiter_saturn}). The challenge for HR 8799 planets is to explain how solids can overcome potential traps and gaps in protoplanetary disk to reach planet  c and e. This warrants future investigations. 

\begin{figure*}[th!]
\begin{tabular}{c}
\includegraphics[width=14.0cm]{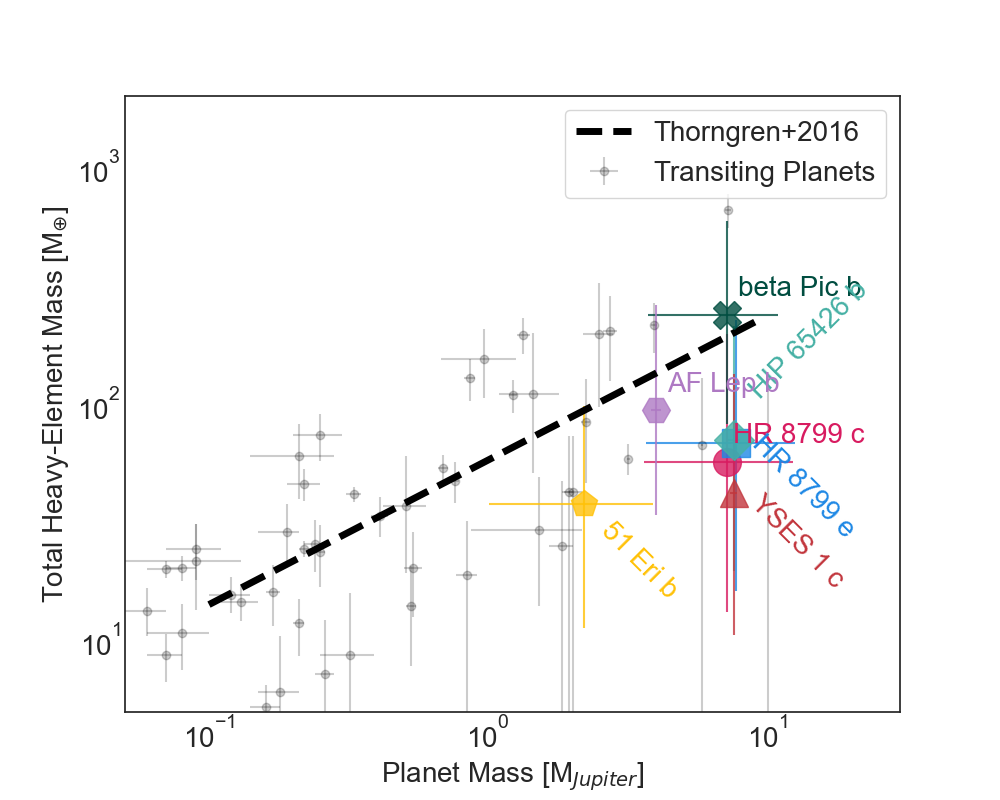}
\end{tabular}
\caption{{\bf{Directly-imaged exoplanets (colored symbols assuming core accretion) roughly follow the same trend as the mass-metallicity relation for transiting exoplanets.}} Dashed line is the mass-metallicity relation reported in~\citet{Thorngren2016} and the grey data points are transiting exoplanets used in deriving the the relation.   }
    \label{fig:transit}
\end{figure*} 

\subsection{Mass-Metallicity Relationship for Transiting and Directly-Imaged Exoplanets}
\label{sec:mass_metallicity}

~\citet{Thorngren2016} study a sample of transiting planets with low stellar irradiation and report a positive mass-metallicity relation, i.e., the mass of heavy element (M$_Z$) scales with planet mass M with a power law of $\sim$0.6, which is consistent the prediction of the CA model. They also concluded that metal-enriched giant planet atmospheres should be the rule because the heavy metal mass for giant planets greatly exceeds the core mass and should therefore enrich the atmospheric metallicity. 

In this work, we significantly increase the sample on the higher mass end by including the seven directly-imaged exoplanets. We show in Fig. \ref{fig:transit} that the mass-metallicity relation roughly holds but starts to show a break at the highest end in the planetary mass regime. While $\beta$ Pic b is located along the linear relationship in the log-log space, other 4 planets fall below the trend and occupy the lower end of total metal mass among all the transiting exoplanets and the directly-imaged exoplanets. This may indicate an upper limit of total metal mass in the planetary mass regime, likely due to an interplay of the time scale of planet formation and the available metals at the time and afterward. 



\section{Summary}
\label{sec:summary}
{{We provide evidence that solid accretion takes place early when the disk is massive and younger than $\sim$2 Myr, which is corroborative with state-of-the-art observations and simulations. In reaching this conclusion, we develop a Bayesian framework (\S \ref{sec:method}) to infer the total mass mass of solid accretion along with other parameters that we marginalize (see Table \ref{tab:prior}). Our results are also compared to previous studies on solar system gas giant planets: Jupiter and Saturn (\S \ref{sec:jupiter_saturn}) as well as individual exoplanet systems (\S \ref{sec:comp}). Model caveats, mainly the strong correlation between the inferred accreted solid mass and the assumed solid-to-gas ratio are discussed in \S \ref{sec:accreted_mass} and \S \ref{sec:beta_pic}. }}


As such, we summarize our major findings here. We analyze a sample of seven directly-imaged exoplanets with measured stellar and planetary chemical abundances (51 Eri b, $\beta$ Pic b, HIP 65426 b, HR 8799 c and e, AF Lep b, and YSES 1 c). We infer a large amount of accreted solids to enrich their atmospheres regardless of formation channels. On average $\gtrapprox$50 M$_\oplus$ solids (ranging from 23.3 to 223.2 M$_\oplus$ for individual systems) are accreted to enrich planet atmospheres. The result, when compared to the available solid content over disk evolution, implies that the solid accretion process can be as early as $\sim${{1-2}} Myrs if available solids decrease drastically with time, but can also last longer if solids growth is more quickly than $\sim${{2}} Myrs.  

Future improvement of this work will come from two angles. First, $\sim$20-30 directly-imaged exoplanets will be characterized by facilities such as JWST, the Keck Planet Imager and Characterizer~\citep[KPIC, ][]{Mawet2018, Jovanovic2019, Delorme2020,Delorme2021}, and VLT/HiRISE~\citep{Vigan2018,Morsy2022}. Our Bayesian framework is ideal to include more objects with data of chemical abundances and elemental ratios. Second, the measurement uncertainty for planetary atmospheric composition (Table \ref{tab:data}) is expected to be much improved by the wavelength coverage of JWST and the high-spectral resolution of KPIC and VLT/HiRISE, the precision of the inference will benefit from the improved measurement uncertainty, although the model in our Bayesian framework will need to be improved accordingly~\citep[e.g., disk evolution and migration, ][]{Molliere2022} to limit the systematic modeling error.

\noindent
{\bf{Acknowledgments}}This work is supported by the National Science Foundation under Grant No. 2143400. The author would like to thank helpful comments and constructive conversations from Rixin Li, Scott Gaudi, and Eric Ford. 

\bibliography{sample63}{}
\bibliographystyle{aasjournal}



\begin{deluxetable}{clllcll}
\tablewidth{0pt}
\tablecaption{Data for Each Exoplanet System \label{tab:data}}
\tablehead{
\colhead{\textbf{}} &
\multicolumn{3}{c}{Star} &
\multicolumn{3}{c}{Planet} \\
\colhead{\textbf{Name}} &
\colhead{\textbf{$\xi_C$}} &
\colhead{\textbf{$\xi_O$}} &
\colhead{\textbf{C/O$_\ast$}} &
\colhead{\textbf{[M/H]$_P$}} &
\colhead{\textbf{C/O$_P$}} &
\colhead{\textbf{$M_P$}}
}


\startdata
51 Eri b & $-3.44\pm0.05^1$ & $-3.18\pm0.05$ & $0.55\pm0.10$ & $0.24\pm0.30$ & $0.38\pm0.09$ & [2.0 ... 4.0]$^2$ \\
$\beta$ Pic b & $-3.57\pm0.05$ & $-3.31\pm0.05$ & $0.55\pm0.10^3$ & $0.75\pm0.10$ & $0.43\pm0.05^4$ & [6.8 ... 11.0]$^5$ \\
HIP 65426 b & $-3.60\pm0.05^6$ & $-3.34\pm0.05$ & $0.55\pm0.10^7$ & $0.05\pm0.25$ & $0.40\pm0.10^8$ & $7.1\pm1.1^9$ \\
HR 8799 c & \multirow{2}{*}{$-3.46\pm0.12$} & \multirow{2}{*}{$-3.19\pm0.14$} & \multirow{2}{*}{$0.54\pm0.10^{10}$} & $0.39\pm0.30^{11}$ & $0.67\pm0.15^{11}$ & [6.5 ... 12.0]$^{12}$ \\
HR 8799 e &  &  &  & $0.34\pm0.25$ & $0.60\pm0.07^{13}$ & [6.5 ... 12.0]$^{12}$ \\
AF Lep b & $-3.84\pm0.31$ & $-3.58\pm0.31$ & $0.55\pm0.10^{14}$ & $0.75\pm0.25$ & $0.55\pm0.10$ & $3.75\pm0.50^{15}$ \\
YSES 1 c & $-3.64\pm0.01$ & $-3.38\pm0.01$ & $0.55\pm0.10^{16}$ & $-0.26\pm0.40$ & $0.36\pm0.15$ & $7.2\pm0.70^{16}$ \\
\hline
Jupiter &  \multirow{2}{*}{$-3.57\pm0.05$} & \multirow{2}{*}{$-3.31\pm0.05$} & \multirow{2}{*}{$0.55\pm0.10^3$}  & $0.61\pm0.10^{17}$ & $0.55\pm0.15^{18}$ & [0.9 ... 1.1]$^{19}$ \\
Saturn &  &  &  & $0.95\pm0.03^{17}$ & $0.55\pm0.15^{18}$ & [0.27 ... 0.33]$^{19}$ \\
\enddata

\tablecomments{1,~\citet{Swastik2021}, assuming solar C/O~\citep{Asplund2009} and scaled from [Fe/H] = 0.13, uncertainty is increased to 0.05; 2,~\citet{BrownSevilla2022}, using Table 5 ``New nominal" column for planet parameters; 3, assuming solar values; 4,~\citet{Nowak2020}, using values in the last row in Table 3, uncertainties are readjusted to 0.10 and 0.05 for metallicity and C/O; 5, lower limit from~\citet{Brandt2021b} and upper limit from~\citet{Dupuy2022}; 6,~\citet{Swastik2021, Bochanski2018}; 7, assuming solar C/O~\citep{Asplund2009}; 8,~\citet{Petrus2021}, using ``K band with continuum" solution in Table 2, uncertainties are readjusted to 0.25 and 0.10 for metallicity and C/O; 9,~\citet{Carter2022}; 10,~\citet{Wang2022}, C/O uncertainty readjusted to 0.10; 11,~\citet{Wang2023}, use {{[O/H]$_P$ and C/O$_P$ in their Table 3 to calculate [M/H]$_P$ based on Eq. \ref{eq:m_h}}}, uncertainty readjusted to 0.30; 12, using the lower mass limit and the dynamical stability upper limit in~\citet{Wang2018b} which encompasses the mass range in~\citet{Brandt2021}; 13,~\citet{Molliere2020}; 14,~\citet{Zhang2023}; 15,~\citet{Balmer2024}; 16,~\citet{Zhang2024}; 17,~\citet{Atreya2018} using C/H as a approximation to M/H; 18, assuming solar C/O but with a larger uncertainty; 19, assuming 10\% mass uncertainty. 
}

\end{deluxetable}

\begin{deluxetable}{clll}
\tablewidth{0pt}
\tablecaption{Adopted Properties for C and O Carriers \label{tab:co_carriers}}
\tablehead{
\colhead{\textbf{Species}} &
\colhead{\textbf{T$_{evap}$}} &
\colhead{\textbf{$n_O$}} &
\colhead{\textbf{$n_C$}} \\
\colhead{\textbf{}} &
\colhead{\textbf{[K]}} &
\colhead{\textbf{[10$^{-4}\times n_H$]}} &
\colhead{\textbf{[10$^{-4}\times n_H$]}} 
}


\startdata
CO & 20 & 1.8 & 1.8 \\
CO$_2$ & 47 & 0.6 & 0.3 \\
H$_2$O & 135 & 0.9 & \nodata \\
Carbon grains & 500 & \nodata & 0.6 \\
Silicate & 1500 & 1.4 & \nodata \\
\enddata


\end{deluxetable}

\begin{deluxetable}{cllll}
\tablewidth{0pt}
\tablecaption{Inference Method \label{tab:input_output}}
\tablehead{
\colhead{\textbf{Input Data}} &
\colhead{\textbf{Input Symbols}} &
\colhead{\textbf{Parameters}} &
\colhead{\textbf{Symbols}} &
\colhead{\textbf{Equation}} 
}


\startdata
Stellar C abundance  & $\xi_C$ & CO Scaling Factor & $\epsilon_{CO}$ & $\xi_C = \log{\sum n_{C_i}\cdot \epsilon_{i}}^{\dagger}$ \\
Stellar O abundance  & $\xi_O$ & CO$_2$ Scaling Factor & $\epsilon_{CO_2}$ & $\xi_O = \log{\sum n_{O_i}\cdot \epsilon_{i}}$ \\
Stellar C/O & C/O$_\ast$ & H$_2$O Scaling Factor & $\epsilon_{H_2O}$ & C/O$_\ast$ = $\frac{{\sum n_{C_i}\cdot \epsilon_{i}}}{{\sum n_{O_i}\cdot \epsilon_{i}}}$ \\
Planet Relative Metallicity & [M/H]$_P$ - [M/H]$_\ast$ & C grain Scaling Factor & $\epsilon_{Cgrains}$ & Eq. \ref{eq:m_h}  \\
Planetary C/O & C/O$_P$ & O grain Scaling Factor & $\epsilon_{Ograins}$ & C/O$_P = \alpha_C/\alpha_O\times C/O_\ast^{\ddagger}$\\
&  & Planet Mass & $M_P$ & $M_g = M_P - M_s^\prime - M_{\rm{core}}$ \\
&  & Accreted Solid Mass & $M_s^\prime$ &  $M_s = M_s^\prime \times \eta$ \\
&  & Solid-to-Gas Ratio & $f_{sg}$ & \\
&  & Disk Temperature & T$_d$ & \\
 &  & Initial Core Mass & $M_{\rm{core}}$ & \\
 &  & Ablation Fraction & $\eta$ & \\
\enddata

\tablecomments{$\dagger$: $n_{C_i}$ can be found in the $n_{C}$ column in Table \ref{tab:co_carriers}; subscript $i$ in $\epsilon_{i}$ can be CO, CO$_2$, H$_2$O, C and O grains. $\ddagger$: definition of $\alpha_C$ and $\alpha_O$ is in Eq. \ref{eq:a_x}.    
}

\end{deluxetable}

\begin{deluxetable}{lcccc}
\tablewidth{0pt}
\tablecaption{Parameters Used in the Bayesian Inference\label{tab:prior}}
\tablehead{
\colhead{\textbf{Parameter}} &
\colhead{\textbf{Unit}} &
\colhead{\textbf{Type}} &
\colhead{\textbf{Lower}} &
\colhead{\textbf{Upper}} \\
\colhead{\textbf{}} &
\colhead{\textbf{}} &
\colhead{\textbf{}} &
\colhead{\textbf{or Mean}} &
\colhead{\textbf{or Std}} 
}

\startdata
CO Scaling Factor ($\epsilon_{CO}$)                 &  \nodata                 &   Uniform        &  0.5        &   2.0       \\
CO$_2$ Scaling Factor ($\epsilon_{CO_2}$)                 &  \nodata                 &   Uniform        &  0.5        &   2.0       \\
H$_2$O Scaling Factor ($\epsilon_{H_2O}$)                 &  \nodata                 &   Uniform        &  0.5        &   2.0       \\
C Grain Scaling Factor ($\epsilon_{Cgrains}$)                 &  \nodata                 &   Uniform        &  0.5        &   2.0       \\
O Grain Scaling Factor ($\epsilon_{Ograins}$)                 &  \nodata                 &   Uniform        &  0.5        &   2.0       \\
Planet Mass ($M_P$) &  $M_{Jupiter}$                &   \multicolumn{3}{c}{see Table \ref{tab:data} for individual planets}       \\
Accreted Solid Mass ($\log M_s^\prime$) &  M$_\oplus$  &   Log-uniform        &   0      &   4       \\
Solid-to-Gas Ratio ($\log f_{sg}$) &  \nodata  &   Log-normal        &   -2.0      &   0.5       \\
Disk Temperature ($\log$T$_d$) &  K   &   Log-uniform      &  1.0       &   2.5      \\
Initial Core Mass ($\log M_{\rm{core}}$) &  M$_\oplus$             &   Log-uniform       &   -1.0     &   2.0      \\
Ablation Fraction ($\eta$) &  \nodata &   Uniform        &   0.5      &   1.0       \\
\enddata



\end{deluxetable}

\begin{deluxetable}{lcccc}
\tablewidth{0pt}
\tablecaption{Mass of Metals in Jupiter, Saturn, and Directly-Imaged Exoplanets\label{tab:jupiter_saturn}}
\tablehead{
\colhead{\textbf{}} &
\colhead{\textbf{~\citet{Guillot1999}}} &
\colhead{\textbf{~\citet{Thorngren2016}$^\dagger$}} &
\colhead{\textbf{$M_s^\prime$ $^{\ddagger}$}} &
\colhead{\textbf{$M_z = M_c + M_s^\prime$ $^{\ddagger}$}}  \\
\colhead{\textbf{}} &
\colhead{\textbf{[M$_\oplus$]}} &
\colhead{\textbf{[M$_\oplus$]}} &
\colhead{\textbf{[M$_\oplus$]}} &
\colhead{\textbf{[M$_\oplus$]}} 
}

\startdata
Jupiter & 10-40 & $37\pm20$ & $ 15.6^{+ 20.1}_{-  8.3}$ & $ 27.6^{+ 25.3}_{- 16.6}$ \\
Saturn & 20-30 & $27\pm8$ & $  9.5^{+  8.8}_{-  5.1}$ & $ 13.9^{+ 11.6}_{-  6.3}$ \\
HR 8799 c & \nodata &\nodata & $ 41.7^{+145.0}_{- 34.5}$& $ 56.9^{+139.8}_{- 43.7}$ \\
HR 8799 e & \nodata &\nodata & $ 51.4^{+159.4}_{- 41.6}$& $ 68.3^{+157.9}_{- 52.2}$\\
51 Eri b & \nodata &\nodata & $ 23.3^{+ 54.6}_{- 17.6}$& $ 37.7^{+ 60.7}_{- 26.4}$\\
$\beta$ Pic b & \nodata &\nodata & $223.2^{+355.4}_{-153.5}$& $237.5^{+357.1}_{-155.4}$\\
HIP 65426 b & \nodata &\nodata & $ 54.1^{+143.3}_{- 42.0}$& $ 70.4^{+143.6}_{- 50.7}$\\
AF Lep b & \nodata &\nodata & $ 79.0^{+176.1}_{- 53.8}$& $ 94.1^{+166.7}_{- 60.1}$\\
YSES 1 c & \nodata &\nodata & $ 26.9^{+ 84.6}_{- 22.1}$& $ 42.0^{+ 91.7}_{- 31.4}$\\
\enddata

\tablecomments{${\dagger}$: use larger uncertainties in Table 2 in~\citet{Thorngren2016} $\ddagger$: assuming formation outside the CO iceline for Jupiter and Saturn; {{16, 50, and 84 percentiles are reported. } }}


\end{deluxetable}


\newpage
\appendix

\section{Diagnostic Plots}
\label{sec:diagnostic}

\begin{figure}[h!]
\epsscale{1.0}
\plotone{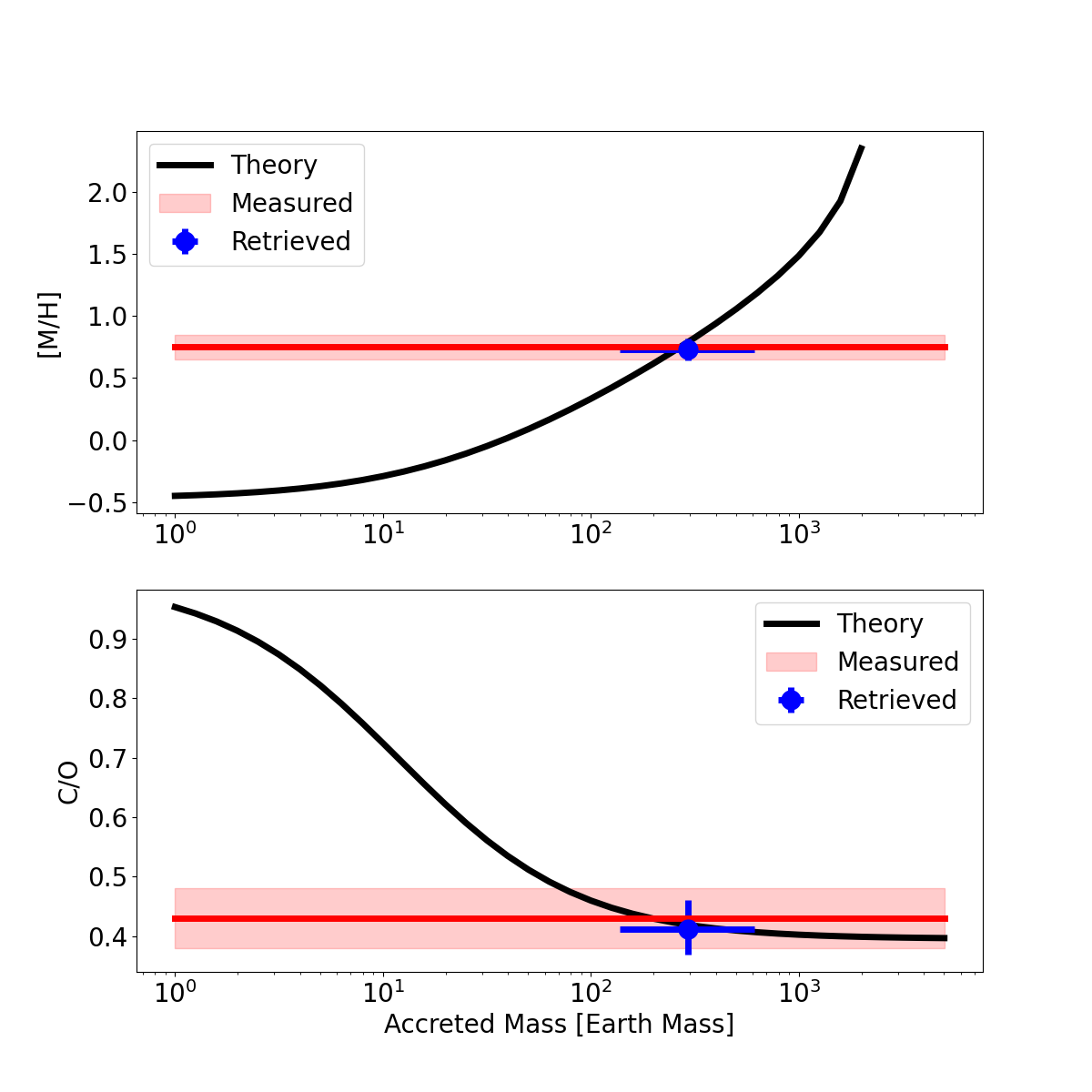}
\caption{Illustration for how $\Delta$[M/H] and C/O change with accretion mass and how the accretion mass (Blue data point with error bars) is determined by comparing model (Black solid curve) to observed values (red solid line and 1-$\sigma$ band).
\label{fig:illus}}
\end{figure} 

\section{Corner Plots}
\label{sec:corner}


\begin{figure}[h!]
\epsscale{1.0}
\plotone{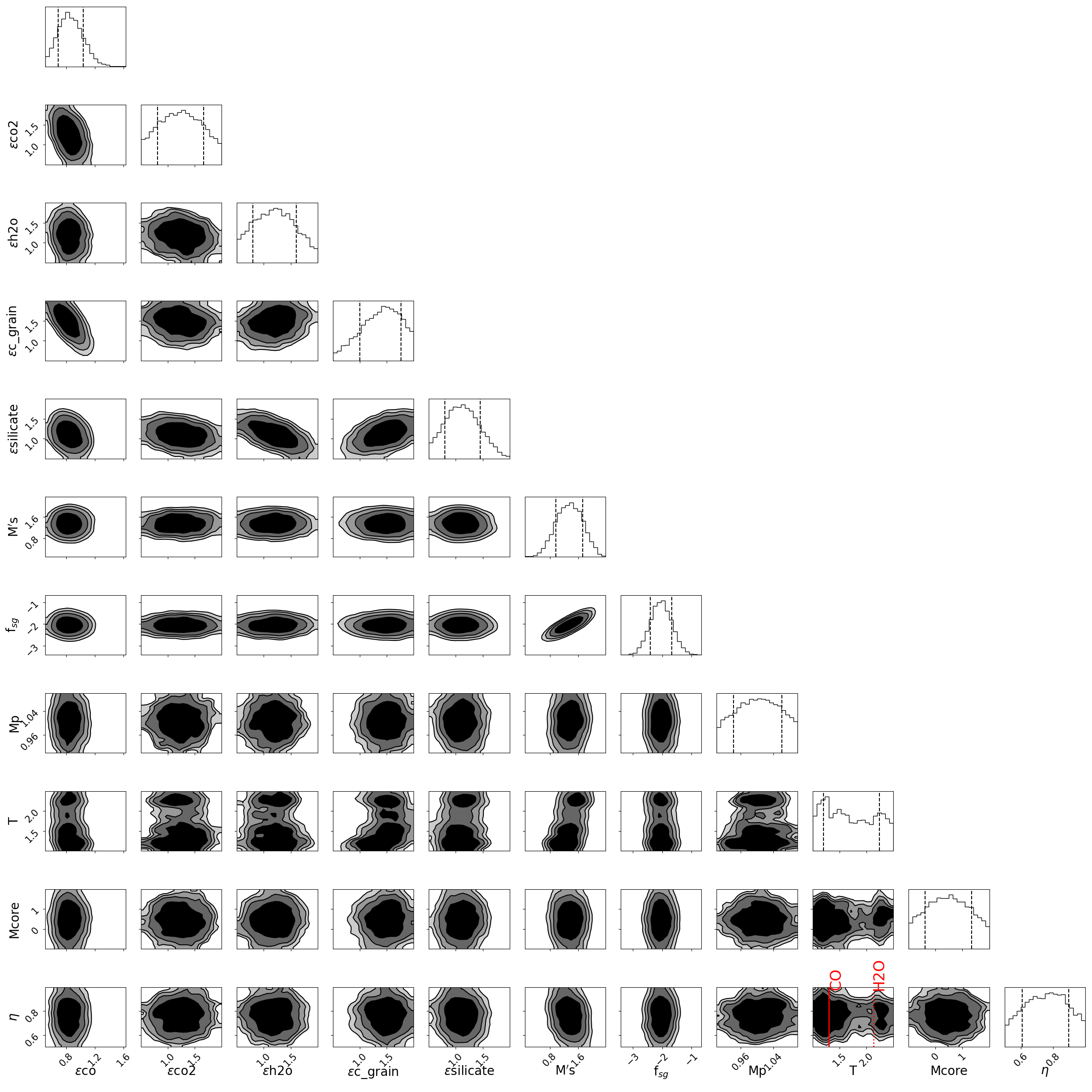}
\caption{Corner plot for Jupiter. {{The corresponding condensation temperatures for CO and H$_2$O are marked by red solid and dotted lines, respectively. }} 
\label{fig:corner_solar_j}}
\end{figure} 

\begin{figure}[h!]
\epsscale{1.0}
\plotone{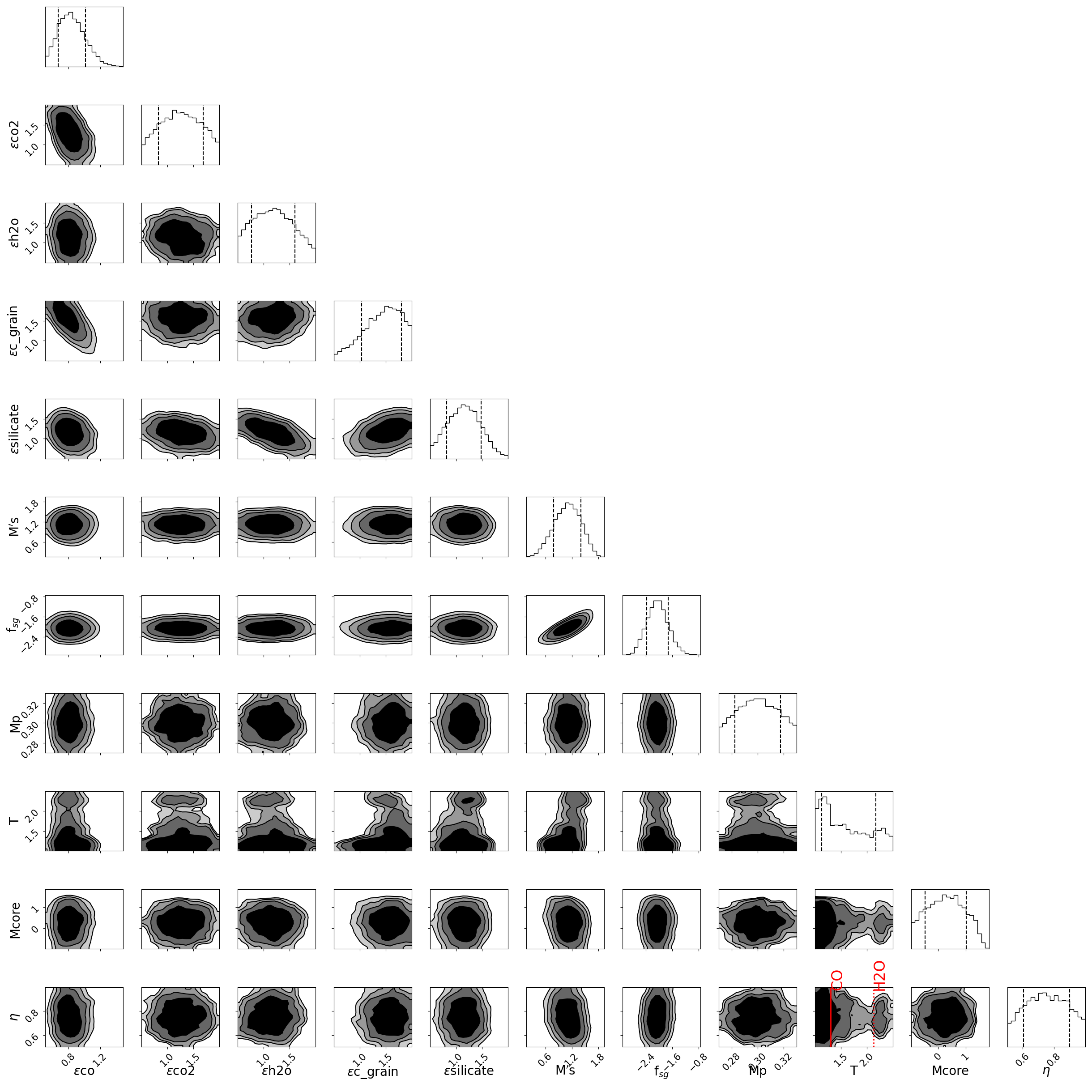}
\caption{Same as Fig. \ref{fig:corner_solar_j} but for Saturn. 
\label{fig:corner_solar_s}}
\end{figure} 

\end{CJK*}
 
\end{document}